# Status of the Tevatron

Simona Rolli
*Tufts University, Medford, MA 02155, USA*

The status of the Tevatron Collider is reviewed and highlights of the rich physics program carried out by the CDF and D0 experiments are presented.

## 1. INTRODUCTION

The Tevatron Collider has been performing remarkably well in the past few years and it is continuing to deliver record luminosity. The machine collides proton and anti-proton beams at an energy in the center of mass of 1.96 TeV, with average peak luminosity of 300E30 $cm^{-2}s^{-1}$. The total delivered luminosity is slightly above 9 $fb^{-1}$. The CDF and D0 experiments have been collecting data with an average efficiency of 90%, while the experiments have enjoyed an annual doubling of the integrated luminosity delivered and recorded. This has led to an avalanche of new results from areas as diverse as QCD, top, searches for new physics and the area of electroweak symmetry breaking with particular focus on direct searches for the Higgs boson. The physics reach of the Tevatron is built on a mountain of measurements that confirm the ability of the Tevatron collaborations to use their detectors to discover new particles. Each measurement is of itself a significant result. Measurements begin with the largest cross section processes, those of B physics, but move on to processes with small branching ratios and backgrounds that are hard to distinguish from the signal. The measurement of $B_s$ oscillations[1] demonstrates the performance of the silicon tracking and vertexing. Discovery of single top production[2], WZ production[3], and evidence for the ZZ production[4] in both leptonic and now hadronic modes[5] provide the final base camp from which the Higgs summit is in sight. Processes such as single top and ZZ act as important messengers heralding the impending arrival of the Higgs. This journey through lower and lower cross section processes represents our approach to provide convincing evidence of these processes, first as discovery then as measurements that constrain the Standard Model.

## 2. TESTS OF THE STANDARD MODEL AND SEARCH FOR NEW PHYSICS

The Standard Model (SM) of particle physics describes the basic constituents of matter and their interactions. Many years of measurements at LEP, SLC and the Tevatron have failed to find any fundamental departure from its predictions, which in some instances have been checked with an accuracy of one part in thousand or better. The global analysis of electroweak observables provide a superb fit to the predictions of this model, which is based on the SU(3)xSU(2)xU(1) gauge theory, but there is still no direct experimental evidence of the underlying dynamics responsible for the electroweak symmetry breaking, the Higgs boson, whose mass, is not predicted by the model. The minimal model contains too many arbitrary parameters for a fundamental theory, which should be simple and elegant in its ability to make predictions. Many questions cannot be answered by the Standard Model, such as the origin and pattern of particle masses, the number of families, the quantization of electric charge, and the number of colors. The Standard Model is clearly a very good description of the physics of elementary particles and the interactions at an energy scale of O(100) GeV/$c^2$ and the actual belief is that it is a low-energy approximation of a more fundamental theory. Theories going beyond the Standard Model should give a solution to all these problems and provide a formalism to answer fundamental questions like the unification of strong, electroweak and gravitational interactions (Grand Unified Theories or GUT's) and the consequent hierarchy problem, which emerges from the introduction of new energy scales, at which electroweak and strong interactions or all four interactions should have the same strength. When testing SM predictions at the Tevatron, the approach has been that of validating the predictions with high accuracy, while at the same time probe new hypotheses by studying key distributions sensitive to new physics input. Jet physics and top pair production are examples of the accuracy reached in cross-section measurements or parameters determination (such as the top quark mass). In addition, the Tevatron has entered a phase of running where SM and beyond SM (BSM) phenomena characterized by very low production cross-sections are finally accessible given the large luminosity collected: CDF and D0 have recently presented evidence of single top as well as multi-boson production for the first time.

The search for the Higgs boson is one of the highest priorities in particle physics programs worldwide. Understanding the mechanism that breaks electroweak symmetry and generates the masses of all known elementary particles is one of the most fundamental problems in particle physics. The Tevatron experiments are making rapid progress in excluding a high mass Higgs boson and could be able to exclude at 3σ confidence level a light-mass one, if it does not exist, with the luminosity projected to be delivered in the next three years. Finally the search for physics beyond the Standard Model (BSM) is a very active area for both the experiments. Two main approaches have been used in a complementary fashion: model-based analysis and signature based studies. In the more traditional model-driven approach, one picks a favorite theoretical model and/or a process, and the best signature is chosen. The selection cuts for acceptances are optimized based on signal MC. The expected background is calculated from data or Monte Carlo and, based on the number of events observed in the data, a discovery is made or the best limit on the new signal is set. In a signature-based approach a specific signature is picked (i.e. dileptons+X) and the data sample is defined in terms of known SM processes. A signal region (blind box) might be defined with cuts kept as loose as possible and the background predictions in the signal region are often extrapolated from control regions. Inconsistencies with the SM predictions will provide indication of possible new physics. As the cuts and acceptances are often calculated independently from a model, different models can be tested against the data sample. It should be noted that the comparison with a specific model, implies calculating optimized acceptances for a specific BSM signal. In signature-based searches, there is no such an optimization.

## 3. RESULTS

### 3.1. QCD Physics

The measurement of the differential inclusive jet cross section at the Fermilab Tevatron probes the highest momentum transfers in particle collisions currently attainable in any accelerator equipment, and thus is potentially sensitive to new physics such as quark substructure. The measurement also provides a direct test of predictions of perturbative quantum chromodynamics (pQCD). The inclusive jet cross section measurements at Tevatron Run II [6, 7, 8, 9] cover up to 600 GeV/c in jet transverse momentum, and range over more than eight orders of magnitude in cross section. Comparisons of the measured cross section with pQCD predictions provide constraints on the parton distribution function(PDF) of the (anti)proton, in particular at high momentum fraction x (x>0.3) where the gluon distribution is poorly constrained[10]. Further constraints on the gluon distribution at high x will contribute to reduced uncertainties on theoretical predictions of many interesting physics processes both for experiments at the Tevatron and for future experiments at the Large Hadron Collider(LHC). One example is top anti-top production at the Tevatron for which the dominant PDF uncertainty arises from the uncertainty in the high-x gluon distribution. In addition, searches for new physics beyond the standard model at high $p_T$ such as quark substructure require precise knowledge of PDFs at high x. Both CDF[11] and D0[12] measure the differential jet cross section using respectively 1.13fb$^{-1}$ and 0.70fb$^{-1}$ of data. Their measurements are in very good agreement with NLO predictions. The experimental precision now exceeds that of the PDF uncertainty, so that such measurements can be used, for the first time, to inform the PDF global fits.

The D0 collaboration has recently determined $\alpha_s$ and its dependence on the momentum transfer using the published measurement of the inclusive jet cross section with the D0 detector at the Fermilab Tevatron Collider in ppbar collisions at √s =1.96 TeV[13]. A combined fit of the data points yields $\alpha_s$ (Mz) = 0.1161+0.0041-0.0048 with $\chi_2$ /Ndf = 17.2/21. The $\alpha_s$ ($p_T$) results support the energy dependence predicted by the renormalization group equation. This is the most precise $\alpha_s$ result obtained at a hadron collider.

The angular distribution of dijets with respect to the hadron beam direction is directly sensitive to the dynamics of the underlying reaction. While in quantum chromodynamics (QCD) this distribution shows small but noticeable deviations from Rutherford scattering, an excess at large angles from the beam axis would be a sign of new physics processes not included in the SM, such as substructure of quarks (quark compositeness), or the existence of additional compactified spatial dimensions (extra dimensions). D0 performs a measurement of the variable $\chi_{dijets}$ = exp( $|y^1 − y^2|$) in ten regions of dijet invariant mass M(jj) , where $y^1$ and $y^2$ are the rapidities of the two jets with highest transverse momentum $p_T$ with respect to the beam axis in an event. For massless 2 → 2 scattering, the variable $\chi$ is related to the polar scattering angle θ∗ in the partonic center-of-mass frame. The choice of this variable is motivated by the fact that Rutherford scattering is independent of χ(dijet), while new physics shows an enhancement at low values of the variable. This is the first measurement of angular distributions of a hard partonic scattering process at energies above 1 TeV in collider-based high energy physics. The normalized χ (dijet) distributions are well described by theory calculations in next-to-leading order in the strong coupling constant and are used to set limits on quark compositeness, ADD large extra dimensions, and TeV$^{-1}$ extra dimensions models[14]. For the TeV$^{-1}$ extra dimensions model this is the first direct search at a collider. For all models considered, this analysis sets the most stringent direct limits to date (see Table in reference [14]).

### 3.2. Top Quark Measurements

After the top quark discovery in 1995 at the Tevatron by both the CDF and D0 collaborations [15, 16], the number of top events available for experimental studies has become more than an order of magnitude larger, thanks to the increasing amount of data delivered by the accelerator. Accurate measurements of the top anti-top production cross section play an important role as tests of QCD NLO calculations and can provide probes towards new physics signals involving non-SM ttbar production mechanisms or decays. Since the CKM element $V_{tb}$ is close to unity and m(top) is large, the SM top quark decays almost exclusively to a W boson and a b quark, so that the top pair production experimental signatures can be classified with respect to the decay modes of the W boson. In 5% of all the top pair decays, when both W 's decay into electrons or muons, we have the so called dilepton channel; in 30% of the decays, when only one W decays into electrons or muons, the lepton plus jets channel. The 14% of the decays involve hadronically decaying taus that are difficult to isolate from the QCD background, while in the 44% of the cases when both W s decay to quarks, we have the so called all hadronic channel. Moreover, the presence of a b-quark in the final state can be exploited as a useful handle to identify the top quark production: b-quarks give rise to jets containing long lived b-hadrons, and those jets can be identified by looking for the presence of tracks in the detector compatible with a secondary decay vertex distinct from the primary interaction point.

CDF provides a combination[17] of measurements of the top quark pair production cross section using a data sample with an integrated luminosity of up to 4.6 fb$^{-1}$. Several measurements carrying different weight in the combination are used and the result is: σ (ttbar) = 7.50 ± 0.48 pb for $m_{top}$ = 172.5 GeV/c$^2$ . The statistical uncertainty is 0.31 pb, the experimental systematic uncertainty is 0.33 pb, Z boson theoretical cross section uncertainty is 0.13 pb, and the luminosity uncertainty is 0.06 pb. The result is in good agreement with the theoretical prediction. There is good agreement among the four measurements with a probability of about 90% to find a less consistent set of measurements. D0 has a similar set of measurements [18].

The top quark mass is a fundamental parameter of the Standard Model (SM). Its precision measurement combined with W boson mass measurement can constrain the mass range of the SM Higgs boson, which is the only unobserved SM particle. For the top mass measurements, two primary techniques have been established. The template method (TM) uses the distributions of variables (templates), which are strongly correlated with the top quark mass and jet energy scale (JES). In the building of a probability, only a few variables (usually less than two) are used, for instance reconstructed top quark mass and dijet mass in the lepton+jets channel. The Matrix Element Method (ME) uses the event probability of being a combination of signals and background. ME method exploits all the information in the event by using a leading order matrix element calculation convoluted with parton distribution function and transfer functions (TFs) making connection between detector response and parton level particle. Because in principle we can use all the information contained in the event, the matrix element technique usually provide better precision than the template method. Both techniques employ a likelihood to compare data to the modeling of signals and background to extract M(top) . D0 has a ME measurement in the lepton+jets channel using 3.6 fb$^{-1}$. D0 employed neural network (NN) based b-tagging to improve signal to background ratio and also reduce the uncertainty in assigning jets to partons. The TF factorizes in contributions from the individual top pair decay products. One can assume that the angles are well measured while their energy and momentum resolutions are determined from MC simulations. D0 estimates TF for four different η regions and for b jets, light jets, and leptons. A W+jets ME is used to estimate background probabilities. In situ JES calibration is performed using dijet mass from hadronically decaying W bosons. D0 measures M(top) = 173.7 ± 1.8 GeV/c$^2$ [19] . CDF also has a ME based measurement in the lepton+jets channel using 4.8 fb$^{-1}$. This analysis integrates over more than 19 variables using a quasi-MC integration technique to account for imperfect assumptions about perfectly measured angles and intermediate particle masses. The TF is parameterized as a function of $\eta$ and $p_T$ separately for b-jets and light jets. This analysis makes use of a neural net to reject not only background contribution but also poorly modeled signal events where the objects in the detector do not match the assumed partons at the matrix element level. In this measurement, the muon acceptance has been increased by using new triggers, giving almost 30% more candidate events with a similar signal to background ratio compared to previous analyses. With in situ JES calibration, CDF measures M(top) = 172.8 ± 1.3 GeV/c$^2$ [20] . This measurement is the most precise top quark mass measurement in the world to date.

With the increased statistical reach of the current data sample, CDF and D0 are measuring several properties of the top-anti top system. In particular, two recent results concerning the forward-backward charge asymmetry seem to point to discrepancies with the SM predictions. CDF measures the forward-backward asymmetry of pair produced top using 1260 fully reconstructed semi-leptonic b-tagged ttbar events in 5.3 fb$^{-1}$ of data collected at CDF[21]. The top rapidity is studied in both the laboratory and the top anti-top rest frames. The parton-level forward-backward asymmetry is found to be to be $A_{lab}$ = 0.150 ± 0.050 ± 0.024 , $A_{tt}$ = 0.158 ± 0.072 ± 0.017 where the first error is statistical and the second is systematic. These results should be compared with the small asymmetries expected in in QCD at NLO, 0.038 ± 0.006 in the lab frame and 0.058 ± 0.009 in the tt frame. The lab frame asymmetry is 3$\sigma$ from null, and more than 2$\sigma$ from the MCFM prediction. Additionally, CDF introduces a simple measurement of the parton level rapidity-dependent asymmetry in two regions of the top anti-top rapidity difference: A( |$\Delta yt$ | < 1.0) = 0.026 ± 0.104 ± 0.055 and A( |$\Delta yt$ | ≥ 1.0) = 0.611 ± 0.210 ± 0.141 to be compared with the MCFM predictions of 0.039 ± 0.006 and 0.123 0.018 for these $\Delta y$ regions respectively. D0 observes a similar discrepancy [22]. They perform the measurement in the lepton + jets final states, with events selected using a b-tagger based on a neural network, and top anti-top candidates fully reconstructed using a kinematic fitter. In 4.3 fb−1 of data they find $A_{lab}$= (0.08 ± 0.04(stat) ± 0.01(syst)), integrated over the acceptance.

### 3.3. Higgs Searches

The Higgs searches at the Tevatron are separated into "high" and "low" mass channels. The high mass channel is characterized by the decay mode into a pair of W bosons, whereas the low mass channels focus on decays to b quark-antiquark pairs or tau pairs. There are four main production mechanisms for the Standard Model Higgs at the Tevatron: gluon-gluon fusion, associated production with a vector boson (W or Z) and vector boson fusion. In the search for the Higgs boson, sequential cut based analyses are not powerful enough to disentangle a very small set of signal events, often buried under a large background. Multivariate analysis techniques do increase the discriminating power of the selection, since they use all available measurements to extract information about the selected event and are widely used in Higgs searches underway at the Tevatron. Both the CDF and D0 collaborations have performed individual combinations of multiple direct searches for the SM Higgs boson[23,24]. The latest searches continue to include more data, adding new channels, and improving analysis techniques compared to previous analyses. The sensitivities of these new combinations improve continuously. The results of both experiments are combined separating the searches into 129 mutually exclusive final states (56 for CDF and 73 for D0 referred to as "analysis sub-channels"[25].

The Higgs boson signal predictions are normalized to the most recent high-order calculations available. The gg→H production cross section is calculated at NNLL in QCD and also includes two-loop electroweak effects, and handling of the running b quark mass. These calculations are refinements of the earlier NNLO calculations of the gg→H production cross section, including electroweak corrections and soft gluon resummation effects. The gg→H production cross section depends strongly on the gluon parton density function, and the accompanying value of $\alpha_s$. The cross sections used are calculated with the MSTW 2008 NNLO PDF set, and the combination includes the larger theoretical uncertainties due to scale variations and PDF variations separately for each jet bin for the gg→H processes. The scale uncertainties are treated as 100% correlated between jet bins and between CDF and D0, and the PDF uncertainties in the cross section are also treated as correlated between jet bins and between CDF and D0. All significant Higgs production modes are included in the high-mass search. For the low-mass searches, we target the WH , ZH , VBF, and ttH  production modes with specific searches, including also those signal components not specifically targeted but which  fall in the acceptance nonetheless. In order to predict the kinematical distributions of Higgs boson signal events, CDF and D0 use the PYTHIA Monte Carlo program, with CTEQ5L and CTEQ6L leading-order (LO) parton distribution functions. The Higgs boson decay branching  ratio predictions are calculated with HDECAY, version 5.3.

All analyses provide binned histograms of the final discriminant variables for the signal and background predictions, itemized separately for each source, and the observed data. The number of channels combined is large, and the number of bins in each channel is large. Therefore, the task of assembling histograms and checking whether the expected and observed limits are consistent with the input predictions and observed data is difficult. We therefore provide histograms that aggregate all channels' signal, background, and data together. In order to preserve most of the sensitivity gain that is achieved by the analyses by binning the data instead of collecting them all together and counting, we aggregate the data and predictions in narrow bins of signal-to-background ratio, s/b. Data with similar s/b may be added together with no loss in sensitivity, assuming similar systematic errors on the predictions. The aggregate histograms do not show the effects of systematic uncertainties, but instead compare the data with the central predictions supplied by each analysis. To gain confidence that the final result does not depend on the details of the statistical formulation, we perform two types of combinations, using Bayesian and Modified Frequentist approaches, which yield limits on the Higgs boson production rate that agree within 10% at each value of m(Higgs) , and within 1% on average. Both methods rely on distributions in the final discriminants, and not just on their single integrated values. Systematic uncertainties enter on the predicted number of signal and background events as well as on the distribution of the discriminants in each analysis ("shape uncertainties"). Both methods use likelihood calculations based on Poisson probabilities.

Systematic uncertainties differ between experiments and analyses, and they affect the rates and shapes of the predicted signal and background in correlated ways. The combined results incorporate the sensitivity of predictions to values of nuisance parameters, and include correlations between rates and shapes, between signals and backgrounds, and between channels within experiments and between experiments. To facilitate comparisons with the Standard model and to accommodate analyses with different degrees of sensitivity, we present our results in terms of the ratio of obtained limits to the SM Higgs boson production cross section, as a function of Higgs boson mass, for test masses for which both experiments have performed dedicated searches in different channels. A value of the combined limit ratio which is less than or equal to one indicates that that particular Higgs boson mass is excluded at the 95% C.L.

The combinations of results[23,24] of each single experiment, as used in this Tevatron combination, yield the following ratios of 95%C.L. observed (expected) limits to the SM cross section: 1.79 (1.90) for CDF and 2.52 (2.36) for D0 at $m_{HIggs}$=115GeV/$c^2$, and 1.13 (1.00) for CDF and 1.02(1.14) for D0 at $m_{Higgs}$=165GeV/$c^2$. The ratios of the 95% C.L. expected and observed limit to the SM cross section are shown in Figure 5 for the Combined CDF and D0 analyses: we found the observed (expected) values of 0.87 (1.24) at $m_{Higgs}$ = 105 GeV/$c^2$ , 1.56 (1.45) at $m_{Higgs}$ = 115 GeV/$c^2$ , 1.28 (1.07)  at $m_{Higgs}$ = 155 GeV/$c^2$ , 0.68 (0.76) at $m_{Higgs}$ = 165 GeV/$c^2$ , 0.95 (1.04) at $m_{Higgs}$ = 175 GeV/$c^2$ and 2.55 (1.61) at  $m_{Higgs}$= 185 GeV/$c^2$ [25].

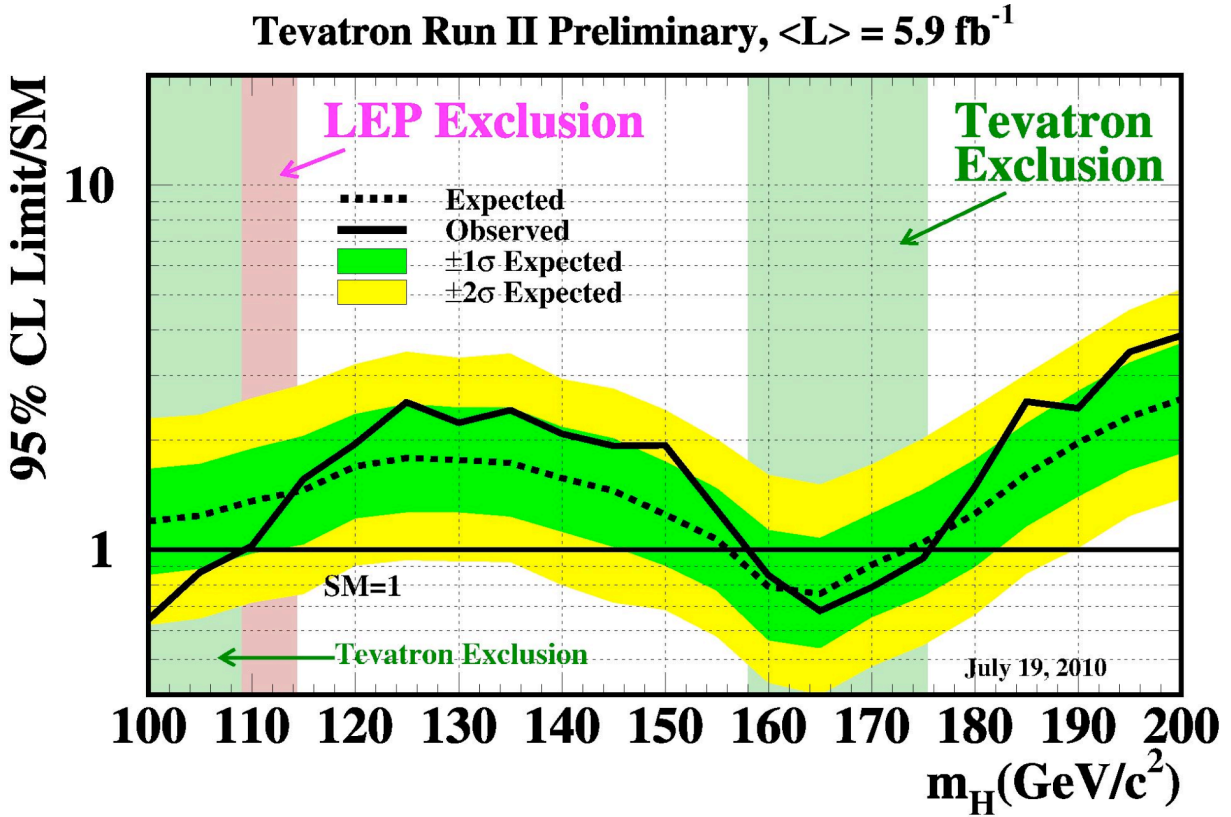

Observed and expected (median, for the background-only hypothesis) 95% C.L. upper limits on the ratios to the SM  cross section, as functions of the Higgs boson mass for the combined CDF and D0 analyses. The limits are expressed as a  multiple of the SM prediction for test masses (every 5 GeV/$c^2$) for which both experiments have performed dedicated searches in different channels. The points are joined by straight lines for better readability. The bands indicate the 68% and 95%  probability regions where the limits can fluctuate, in the absence of signal. The limits displayed in this figure are obtained with  the Bayesian calculation.

### 3.4.  Search for New Physics

Many searches for physics beyond the Standard Model are performed at the Tevatron , from simple mass bump hunts to more refined analyses, characterized by complex final state signatures.  Several of the most recent results are described in these same proceedings [26,27] and we are not going to discuss them further. Here we are only going to report on a recent measurement by the D0 experiment of the charge asymmetry A of like-sign dimuon events in 6.1 fb$^{-1}$ of proton anti-proton collisions[28]. From A, the like-sign dimuon charge asymmetry in semileptonic b-hadron decays is extracted: $A_b^{sl} = -0.00957 \pm 0.00251$ (stat) $\pm 0.00146$ (syst). This result differs by 3.2 standard deviations from the standard model prediction $A_b^{sl}(SM) = (-2.3^{+0.5}_{-0.6}) \times 10^{-4}$ and provides first evidence of anomalous CP-violation in the mixing of neutral B mesons.

The Tevatron is in a unique position to study possible effects of C P violation, in particular through the study of charge asymmetries in generic final states, given that the initial state is CP-symmetric. The high center-of-mass energy provides access to mass states beyond the reach of the B-factories. The periodic reversal of the D0 solenoid and toroid polarities also results in a cancellation at the first order of most detector-related asymmetries. The like-sing dimuon asymmetry is defined as $A = (N^{++} - N^{--})/(N^{++} + N^{--})$ where $N^{++}$ and $N^{--}$ represent, respectively, the number of events in which the two muons of highest transverse momentum satisfying the kinematic selections have the same positive or negative charge. After removing the contributions from backgrounds and from residual detector D0 observes a net asymmetry that is significantly different from zero. The result is interpreted by assuming that the only source of this asymmetry is the mixing of neutral B mesons that decay semileptonically. Therefore a measurement of the asymmetry $A_{sl}^b$ defined as $A_{sl}^b = (N_b^{++} - N_b^{--})/(N_b^{++} + N_b^{--})$ where $N_b^{++}$ and $N_b^{--}$ represent the number of events containing two b hadrons decaying semileptonically and producing two positive or two negative muons, respectively, is obtained. $A_b^{sl}$ is extracted from two observables. The first is the like-sign dimuon charge asymmetry and the second observable is the inclusive muon charge asymmetry a defined as $a = (n^+ - n^-)/(n^+ + n^-)$ where $n^+$ and $n^-$ correspond to the number of detected positive and negative muons, respectively. At the Fermilab Tevatron collider, b quarks are produced mainly in b anti-b pairs. The signal for the asymmetry A is composed of like-sign dimuon events, with one muon arising from direct semileptonic b-hadron decay $b \to \mu^- X$, and the other muon resulting from $B_q^0$ anti-$B_q^0$ oscillation, followed by the direct semileptonic decay of the B-hadron. Consequently the second muon has the "wrong" sign due to the B-hadrons mixing. For the asymmetry a, the signal comes from mixing, followed by the semileptonic decay of the B-hadron. The main backgrounds for these measurements arise from events with at least one muon from kaon or pion decay, or from the sequential decay of b quarks $b \to c \to \mu^+ X$. For the asymmetry a, there is an additional background from direct production of c-quarks followed by their semileptonic decays. A and a are measured starting from a dimuon data sample and an inclusive muon sample respectively. Background processes and detector asymmetries contribute to these asymmetries. These contributions are measured directly in data and used to correct the asymmetries. After applying these corrections, the only expected source of residual asymmetry in both the inclusive muon and dimuon samples is from the asymmetry $A_b^{sl}$. Simulations are used to relate the residual asymmetries to the asymmetry $A_b^{sl}$, and to obtain two independent measurements of $A_b^{sl}$. These measurements are combined to take advantage of the correlated contributions from backgrounds, and to reduce the total uncertainties in the determination of $A_b^{sl}$.

## 4. CONCLUSIONS

The Tevatron Collider continues to produce high quality data from proton anti-proton collisions and the CDF and D0 experiments are producing many interesting results, ranging from Standard Model tests to search for new physics. A wide range of physics processes are in fact studied: from QCD jet production to the most precised measurement of $\alpha_s$ at hadron colliders. Top quark production and properties are studied in many way, and the top quark mass is now known at a precision less than 1%. Together with the best measurement of the W-boson mass, this parameter is a critical input to electroweak fits and constraints the predicted Higgs mass. New tantalizing indications of new physics have been recently evidenced by a D0 measurement of anomalous dimuon charge asymmetry in B-meson systems. Due to the large luminosity collected, small cross-section phenomena are now accessible for study, such as single top electroweak production, diboson and Higgs production. Both the experiments are very actively searching for the Higgs boson, as evidence for it in the mass range favored by current theoretical fits of electroweak data is within reach at the Tevatron especially if the machine will continue to run past 2011.